\documentstyle[aps,multicol,epsf,citesort]{revtex}
\draft
\newcommand{\beq}{\begin{equation}}
\newcommand{\eeq}{\end{equation}}
\newcommand{\bdis}{\begin{displaymath}}
\newcommand{\edis}{\end{displaymath}}
\newcommand{\bea}{\begin{eqnarray}}
\newcommand{\eea}{\end{eqnarray}}
\newcommand{\barr}{\begin{array}}
\newcommand{\earr}{\end{array}}

\begin{document}
\title
{Dispersity-Driven Melting Transition in Two Dimensional Solids}  

\author{M. Reza Sadr-Lahijany$^1$, Purusattam Ray$^2$
 and H. Eugene Stanley$^1$ }

\address{$^1$Center for Polymer Studies and Department of Physics,
        Boston University, Boston, Massachusetts 02215\\
        $^2$The Institute of Mathematical Sciences,
        CIT Campus, Chennai - 600 113, India }

\date{May 15, 1997 }

\maketitle

\begin{abstract}
We perform extensive simulations of $10^4$ Lennard-Jones particles to
study the effect of particle size dispersity on the thermodynamic
stability of two-dimensional solids.  We find a novel phase diagram in
the dispersity-density parameter space. We observe that for large values
of the density there is a threshold value of the size dispersity above
which the solid melts to a liquid along a line of first order phase
transitions. For smaller values of density, our results are consistent
with the presence of an intermediate hexatic phase. Further, these
findings support the possibility of a multicritical point in the
dispersity-density parameter space.

%33757.00
\end{abstract}
\date{srs.tex ~~~ May 15, 1997 ~~~ draft}
\pacs{PACS numbers: 64.70.Dv,64.60.Cn,02.70.Ns,61.20.Ja,05.70.Fh}

\begin{multicols}{2}

Recently there has been considerable interest in what happens to the
liquid-solid transition in a system if the constituent particles are not
all identical but have different sizes.  The question was first raised
in the context of colloidal solutions\cite{Dickinson}, and subsequently
addressed for other systems \cite{Barrat-Hansen,Pusey,Ito}. These
studies mainly focused on the effect of size dispersity $\Delta$ on the
$P-\rho$ equation of state, where $P$ and $\rho$ denote pressure and
density. On increasing $\Delta$ from zero, the density discontinuity at
the transition decreases, eventually vanishing at a critical value
$\Delta=\Delta_c$ above which there is no liquid-solid density
discontinuity. This remarkable phenomenon--- similar to the effect of
temperature $T$ on the conventional liquid-gas phase
transition\cite{footnote}---occurs in both two and three dimensions, and
for various forms of interaction potentials and size
distributions\cite{Ito}.

These seminal studies leave some questions unanswered. First, what are
the structures of the phases? Second, can one pass continuously from
solid to liquid ``around the critical point'' at $\Delta_c$, just as one
can pass continuously from liquid to gas ``around the critical point''
at $T_c$? A ``yes'' answer would not be consistent with the common
picture of melting as a first order phase transition (which cannot have
a critical point because of symmetry mismatch of the two phases
\cite{Landau-Stanley}). A ``no'' answer would lead to a natural third
question: In the $\Delta-\rho$ parameter space, what is the location and
nature of the phase boundary between crystalline and liquid phases? The
third question has not gone unnoticed---indeed, Ref. \cite{Bocquet}
simulates a binary mixture of $108$ ``soft'' disks, and shows that upon
increasing $\Delta$ the crystal undergoes a transition to an amorphous
solid at a threshold dispersity $\Delta_{th}$, suggesting that the
transition is of first order.

Here we address all three questions by simulating a relatively large
system comprised of $N=10^4$ Lennard-Jones
particles of two different radii in a square box of edge $L_0$ with
periodic boundary condition. With each particle $i$, we associate a
size parameter $\sigma_i$, and define the distance scale for the
interaction between particles $i$ and $j$ to be
$\sigma_{ij}\equiv\sigma_i + \sigma_j$. We assign to half the particles
the value $\sigma_i=\sigma_0(1+\Delta)$, and to the other half the value
$\sigma_i=\sigma_0(1-\Delta)$. If particles $i$ and $j$ are at a
distance $r_{ij}$ smaller than a cutoff distance $r_c$, they interact
via a ``shifted-force Lennard-Jones'' potential \cite{Allen-Tildesley}
$\Phi_{ij} = 4 \epsilon \left[\left(\sigma_{ij}/r_{ij}\right)^{12} -
\left(\sigma_{ij}/r_{ij}\right)^{6}\right] + f(r_{ij})$.
%\Phi_{ij} = 0 \qquad \qquad \qquad r_{ij}>r_c
Here $f(r_{ij})$ is a linear function whose coefficients are
chosen such that $\Phi_{ij}$ and its gradient, the force, continuously
vanish at $r_{ij}=r_c$.  Since $\Phi_{ij}$ takes its minimum value
at $r_{ij} = R_{ij}\equiv2^{1/6}\times \sigma_{ij}$, we consider this
equilibrium distance to be the sum of the radii of the two particles $i$
and $j$, $R_{ij}=R_i+R_j$, so the radius of particle $i$ is $R_i =
2^{1/6}\times \sigma_i$ and the average radius is
$2^{1/6}\times\sigma_0$.

We perform molecular dynamics (MD) simulations using the velocity
Verlet integrator method \cite{Allen-Tildesley}. We record the results
in reduced units in which $\sigma_0$ is $0.5$, and the Lennard-Jones
energy scale $\epsilon$, the particle mass, and Boltzmann constant are
all unity. In these units, we choose $r_c=2.5$ and the length of each MD
time step $\delta t=0.01$. The system is first thermalized at $T=1$,
using the Berendsen rescaling method \cite{Allen-Tildesley}, for a
period of length $\tau$; typically $\tau=(5\times10^4)\delta t$. Then we
run the system for an additional period $\tau$ as a constant NVE system
(micro-canonical ensemble). We continuously calculate $P,T$ and energy
$E$, and we consider the system to be in equilibrium only when the
fluctuations of all three quantities are less than $1 \%$ of their
average values. The thermalization time $\tau$ is chosen to be more than
the time it takes for the system to equilibrate.

We define the size dispersity to be the ratio of the size distribution
variance to its average \cite{Pusey}, which equals $\Delta$ in our
model, and we define $\rho\equiv\sum_{i=1}^N (\pi R_i^2)/L_0^2$, the
ratio of the total area assigned to the disks to the system area. For
each value of $\Delta$, we start by placing the $10^4$ particles
randomly on the sites of a square lattice of edge $L_0\approx150$;
higher density states are obtained by gradually compressing the system
by reducing $L_0$. Typically the starting density is $\rho=0.7$, and we
increase $\rho$ to $1.05$ through approximately $10$ intermediate
densities, equilibrating the system at each\cite{simlation}.

We present our results for the state points with $T=1$, $\rho=0.90-1.05$
and $\Delta=0-0.12$. At these densities, the $\Delta=0$ system is a
2D-solid with a triangular order, but at large $\Delta$ the system
becomes disordered and a liquid. By probing the translational and
orientational order, we determine the phase of each state point and we
locate the transition between the two phases. To study translational
order, we calculate the total pair correlation function $g(r)$, as well
as the partial functions $g_{11}(r)$, $g_{22}(r)$ and $g_{12}(r)$
\cite{Allen-Tildesley}. Here $g(r)$ is the probability distribution of
finding two particles at a distance $r$, and $g_{ij}(r)$ is the same for
an $(i,j)$ pair ($i=1$ stands for small and $i=2$ for large
particles). We find that all three $g_{ij}(r)$ display behavior similar
to $g(r)$, indicating that the system maintains its
substitutionally-disordered configuration and does not tend toward
de-mixing.

\begin{figure}[htb]
\narrowtext
\centerline{
\hbox  {
        \vspace*{0.5cm}
        \epsfxsize=4.3cm
        \epsfbox{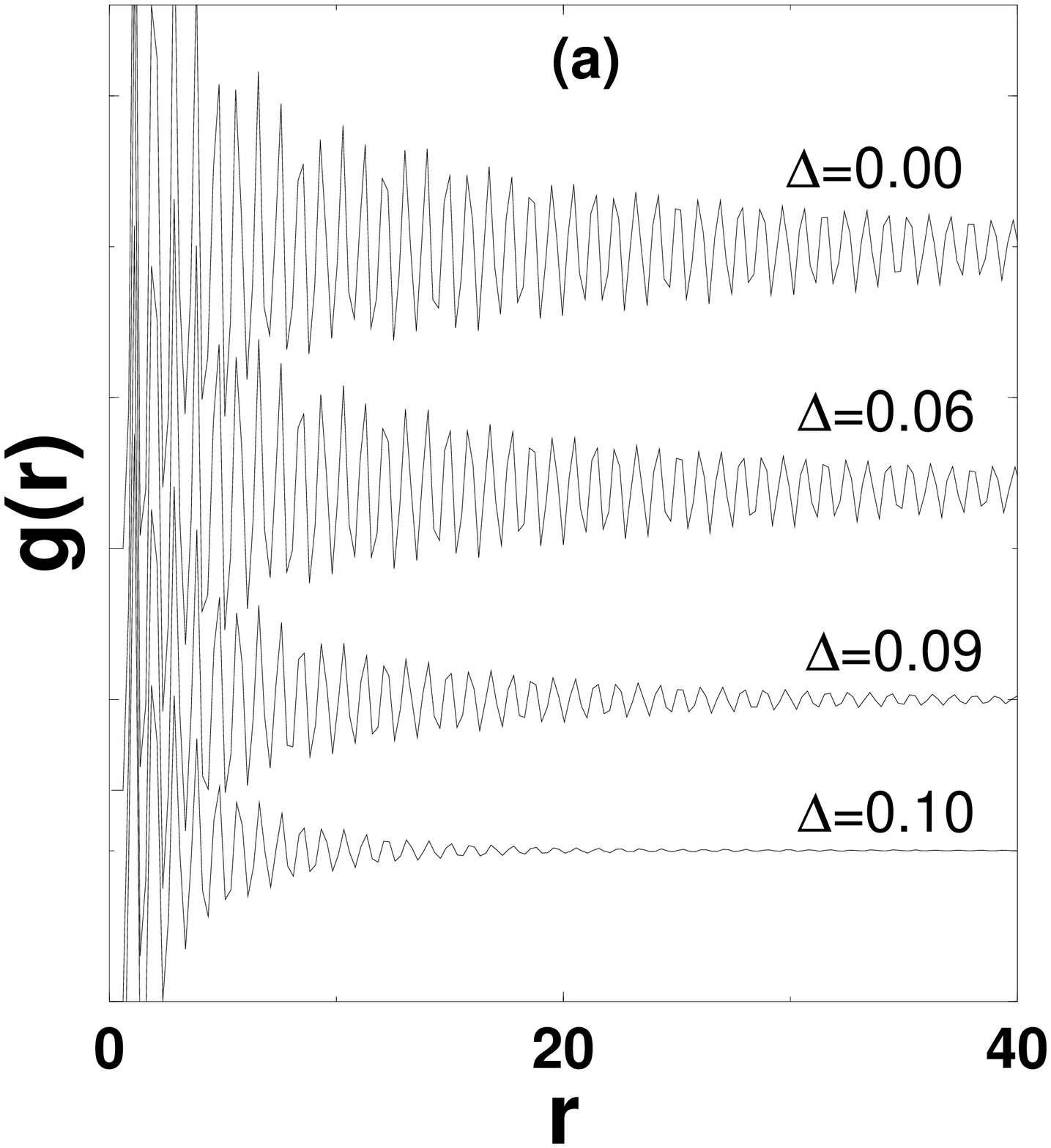}
       \hspace*{0.3cm}
        \epsfxsize=4.3cm
        \epsfbox{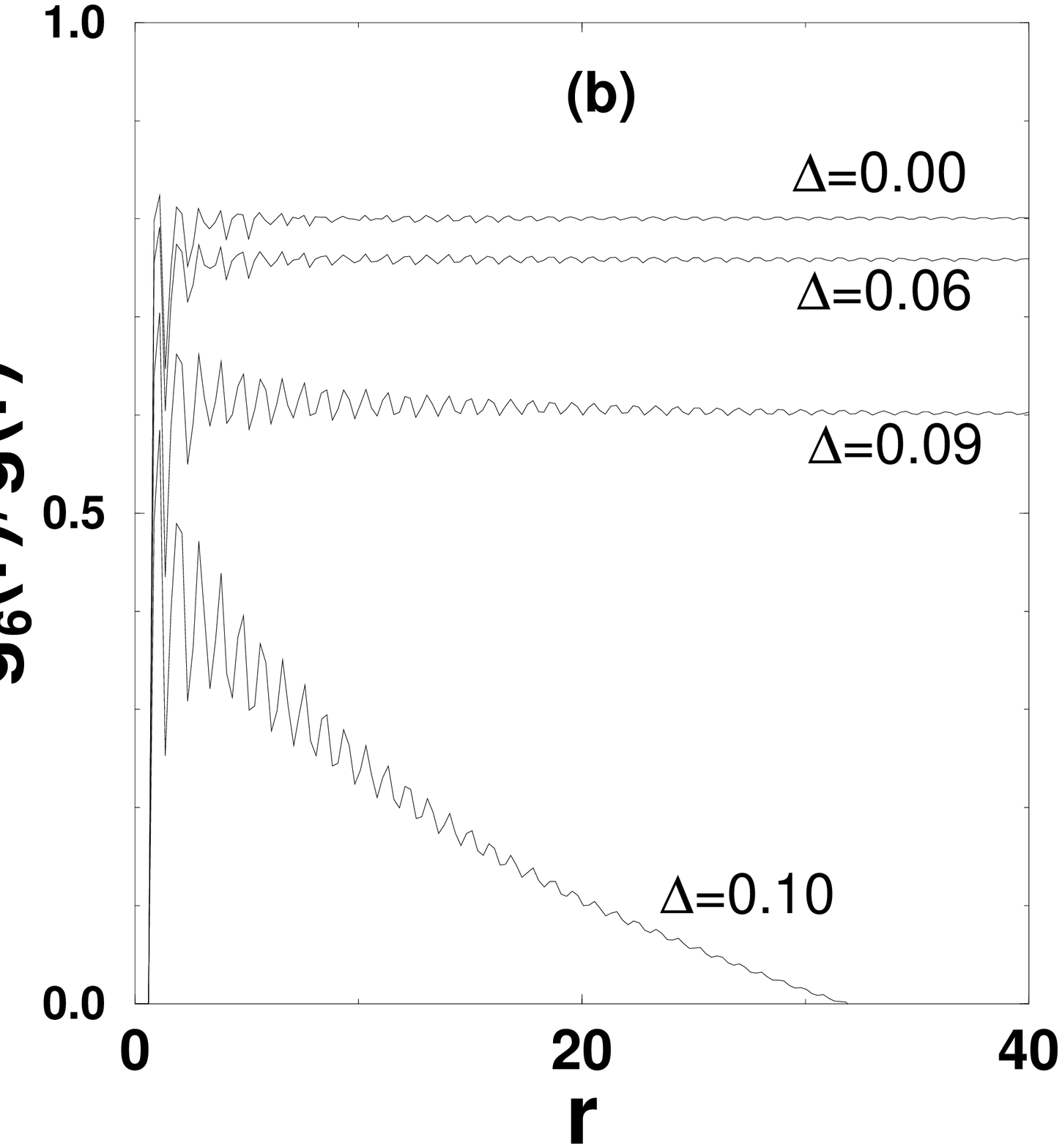}
        \vspace*{0.5cm}
       }
          }     
\caption{Effect of dispersity $\Delta$ on translational and
orientational order at $\rho=1.0$. (a)Total pair correlation function
$g(r)$ as a function of distance $r$.  All curves oscillate around the
value $g(r)=1$, so we have separated them to facilitate comparison.  We
find that a  transition from solid to liquid structure occurs on
increasing the dispersity between $\Delta=0.09$ and
$\Delta=0.10$. (b) Normalized orientational correlation function versus
$r$. The change between $\Delta=0.09$ and $\Delta=0.10$
corresponds to the transition from an orientationally long-range
correlated solid to a short-range correlated liquid. Both sets of curves
show that for $\rho=1.0$ the solid-liquid transition occurs at a value
of dispersity between $0.09$ and $0.10$ which is consistent with the observations based
on Figs. $2-4$.}
\label{figg}
\end{figure}

In Fig.~\ref{figg}(a) we show the effect of tuning $\Delta$ on
translational order. We observe that the monodisperse ($\Delta=0$)
system shows the quasi-long-range translational order expected for a
2D-solid\cite{Nelson}, characterized by a power-law decay of the
envelope of $g(r)$ and the persistence of the solid structure
periodicity up to very large distances. For $\Delta<\Delta_{th}(\rho)$,
where $\Delta_{th}(\rho)$ is the threshold value at fixed $\rho$, the
solid maintains this quasi-long-range order, although the decay exponent
appears to increase somewhat with $\Delta$. For $\Delta>\Delta_{th}(\rho)$, we
observe a qualitative change in the structure: the quasi-long-range
translational order disappears, and is replaced by an exponential decay
of the envelope of $g(r)$, which at very long distances shows the
uniform distribution of a structureless liquid. We observe this behavior
for all densities between $\rho=0.96$ and $\rho=1.05$, and find that
$0.09<\Delta_{th}(\rho)<0.10$ for all $\rho$.

Next we study the local bond orientational order by calculating 
for each particle $j$ the sixfold orientational order parameter\cite{Nelson-Halperin}

\begin{equation}
\label{eqpsi6}
(\psi)_j\equiv{1\over z}\sum_{k=0}^z e^{i6\theta_{jk}}.
\end{equation}

\noindent The sum runs over all $z$ nearest neighbors $k$ of
$j$, and $\theta_{jk}$ is
the angle of the bond joining particles $j$ and $k$ with respect to a
fixed axis. We identify the nearest neighbors as the particles that are
closer than the location of the first minimum of $g(r)$. The
modulus of $(\psi)_j$ will be unity if the neighbors form a perfect
hexagon around $j$, which occurs for all particles in a 
triangular lattice, the close-packed configuration of a 2D-solid. For a
distorted hexagon or a different polygon, $|(\psi)_j|<1$---
e.g. for a liquid, the distribution of $|(\psi)_j|$ centers around $0.5$
\cite{Clark}.

We define the continuous  order parameter field $\psi(\mbox{\bf r})$ as the
value of $(\psi)_j$ if the position of particle $j$ is
$\mbox{\bf r}_j=\mbox{\bf r}$, and we calculate the orientational
correlation function\cite{McTague}

\begin{equation}
\label{eqg6}
g_6(|\mbox{\bf r} -\mbox{\bf r}_0|)\equiv\langle\psi(\mbox{\bf r})\psi(\mbox{\bf r}_0)\rangle,
\end{equation}

\noindent where $\langle...\rangle$ denotes an average over $\mbox{\bf
r},\mbox{\bf r}_0$ and time. Fig.~\ref{figg}(b) shows that if $\Delta$
is small and $\rho$ is large, the system displays the long-range
orientational order of a solid in that $\lim_{r\rightarrow
\infty}g_6(r)\neq0$. Noteworthy is that for each value of $\rho$,
orientational order disappears upon a small increase in dispersity near
$\Delta_{th}(\rho)$. For $\Delta>\Delta_{th}(\rho)$, $g_6(r)$ appears to
decay exponentially, which identifies the system as a liquid. Similar
plots for other large values of $\rho$ suggest that near $\rho=1.0$,
there is a first order phase transition from solid to liquid, driven by
an increase in $\Delta$. This observation is in agreement with the
results of Ref.~\cite{Bocquet}. Fig.~\ref{figg} shows that the small
dispersity system has the {\em ordered} structure of a solid while the
large dispersity system has the {\em disordered} structure of a
liquid---providing an answer to the first of the three questions.

Since identifying phases relies on the behavior of the system in the
thermodynamic limit, we apply finite size scaling to the moments of the
orientational order parameter $\psi$, which is the average value of
$\psi(\mbox{\bf r})$. We use standard techniques originally developed
for the Ising model \cite{Binder-ising}, and recently applied to the 2D
melting transition \cite{Weber-Binder,Bagchi-Andersen}. In order to
calculate the moments of $\psi$ at a scale ${b}\equiv L_0/M$, we divide
the system into $M^2$ blocks of edge ${b}$ and we define
$\psi_{b}$ for each block as the absolute value of the average of
$\psi(\mbox{\bf r})$ over the block. Then we find the moments of
$\psi_{b}$ by averaging over all blocks and all configurations of
the system after equilibration\cite{Weber-Binder}.

\begin{figure}[htb]
\narrowtext
\centerline{
        \vspace*{0.5cm}
        \epsfxsize=4.8cm
        \epsfbox{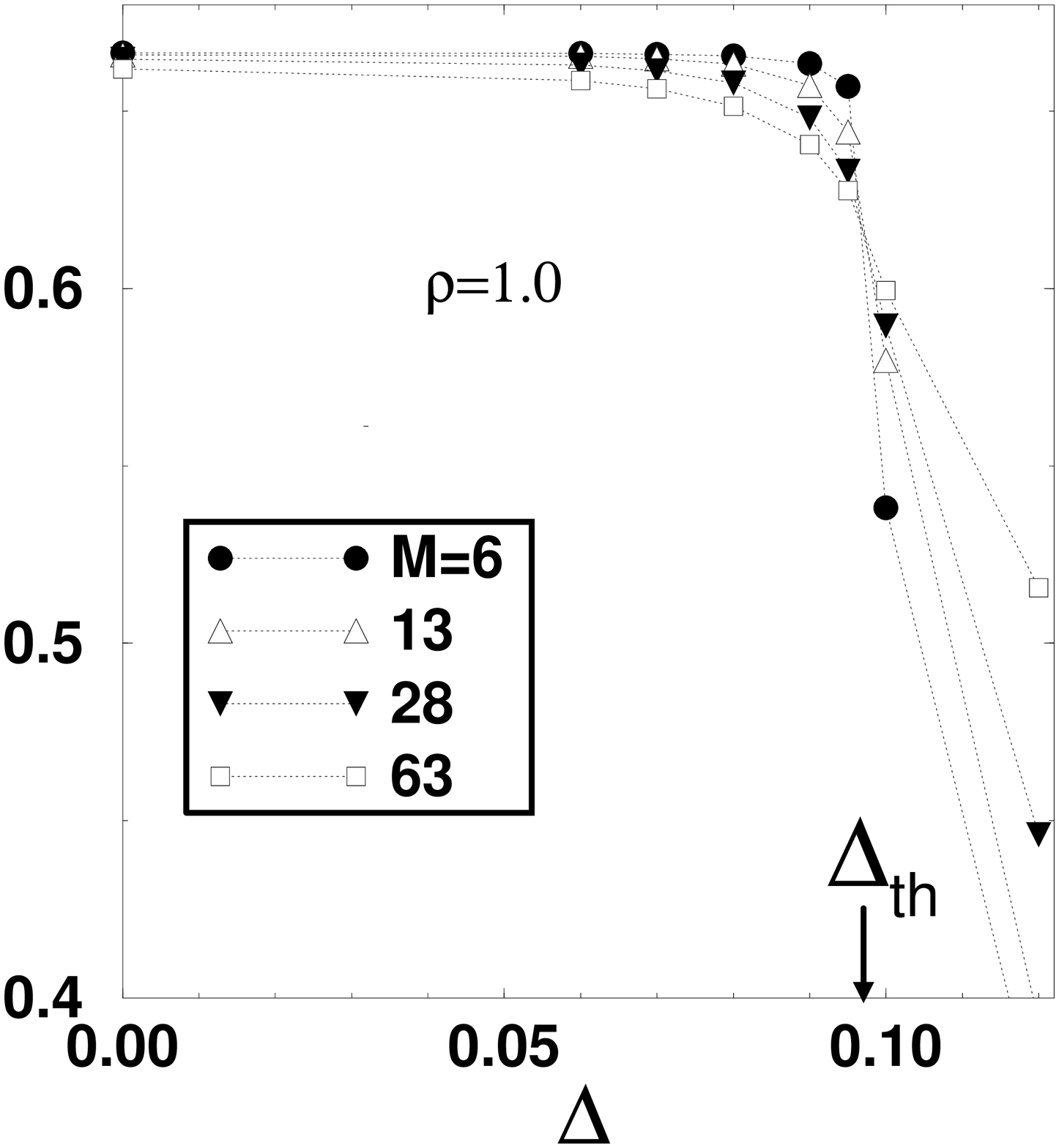}
        \vspace*{0.5cm}
        }
\caption{Cumulant $U_{b}$ of the bond orientational order parameter
$\psi$ as a function of dispersity $\Delta$ for $\rho=1.0$. Different
curves correspond to different scales ${b}$, where ${b}\equiv L_0/M$ is
the block size, so smaller $M$ corresponds to larger scale. The
dotted lines connecting the data points are guides to the eye. We
identify the threshold value of $\Delta$ to be the point where all the
curves for different scales intersect.}
\label{figul}
\end{figure}

\begin{figure}[htb]
\narrowtext
\centerline{
\hbox  {
        \vspace*{0.5cm}
        \epsfxsize=5.5cm
        \epsfbox{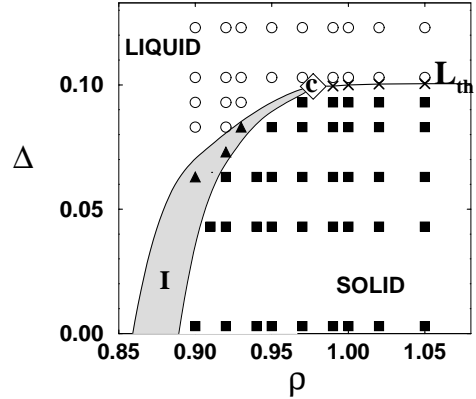}
        \vspace*{0.5cm}
       }
          }     
\caption{ Phase diagram in the
 $\Delta-\rho$ (dispersity-density) parameter space. Squares represent solid points and
circles liquid points. The threshold line $L_{th}$ connects crosses,
which are the first order phase
transition points derived from the cumulant analysis. The
triangles are the points of the intermediate ($I$) phase, showing a
hexatic behavior. The large diamond marks the multicritical point $C$.}
\label{phasediagram}
\end{figure}

To explore the precise location of the phase
transition, we calculate the cumulant \cite{Weber-Binder}
\begin{equation}
\label{equl}
U_{b}\equiv1-{{\langle\psi_{b}^4\rangle}\over{3\langle\psi_{b}^2\rangle^2}}.
\end{equation}
For a completely {\em ordered} (solid) system, $U_{b}=2/3$ in the
thermodynamic limit, while for a {\em disordered} (liquid) system
$U_{b}\rightarrow 0$. For an infinite system, $U_{b}$ jumps between
these two limiting values at the phase transition point and for finite
systems, this jump becomes rounded. Still, one can determine the
location of a phase transition by finding the point at which the $U_{b}$
curves for different system sizes intersect \cite{Weber-Binder}. In
Fig.~\ref{figul}, we plot $U_{b}$ versus $\Delta$, for different scales
${b}$ at a fixed $\rho$. We find the transition from the value $2/3$ to
lower values upon passing through the phase transition. Moreover, we
estimate $\Delta_{th}(\rho)$ from the crossing point of the curves.  In
the phase diagram of Fig.~\ref{phasediagram}, the line $L_{th}$ is the
locus of all such threshold points separating solid and liquid phases
and shows that for $\rho>0.96$, $\Delta_{th}(\rho)\approx0.097$ is
essentially independent of $\rho$.

\begin{figure}[htb]
\narrowtext
\centerline{
\hbox  {
        \vspace*{0.5cm}
        \hspace*{0.2cm}
        \epsfxsize=4.3cm
        \epsfbox{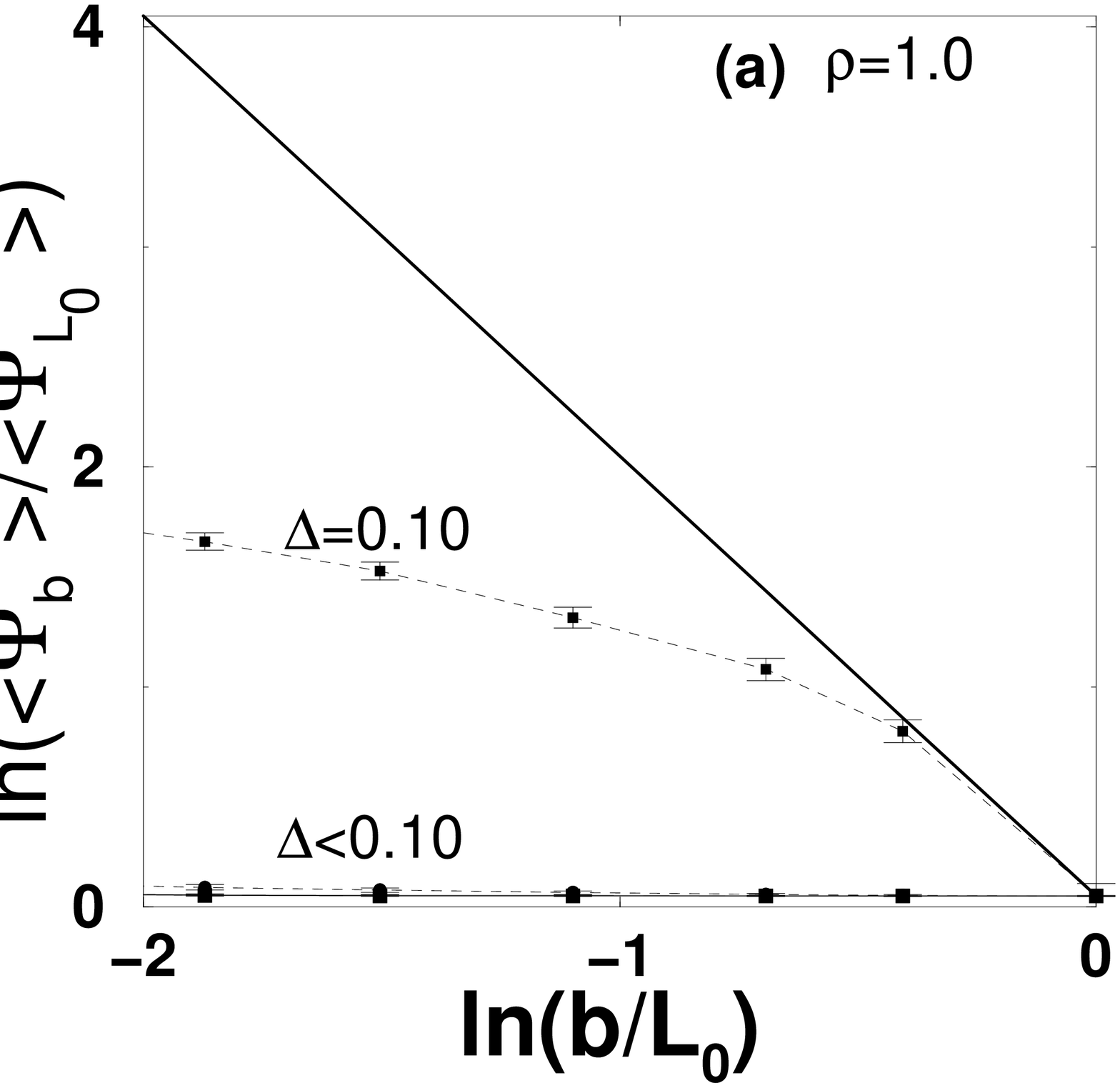}
        \hspace*{0cm}
        \epsfxsize=4.3cm
        \epsfbox{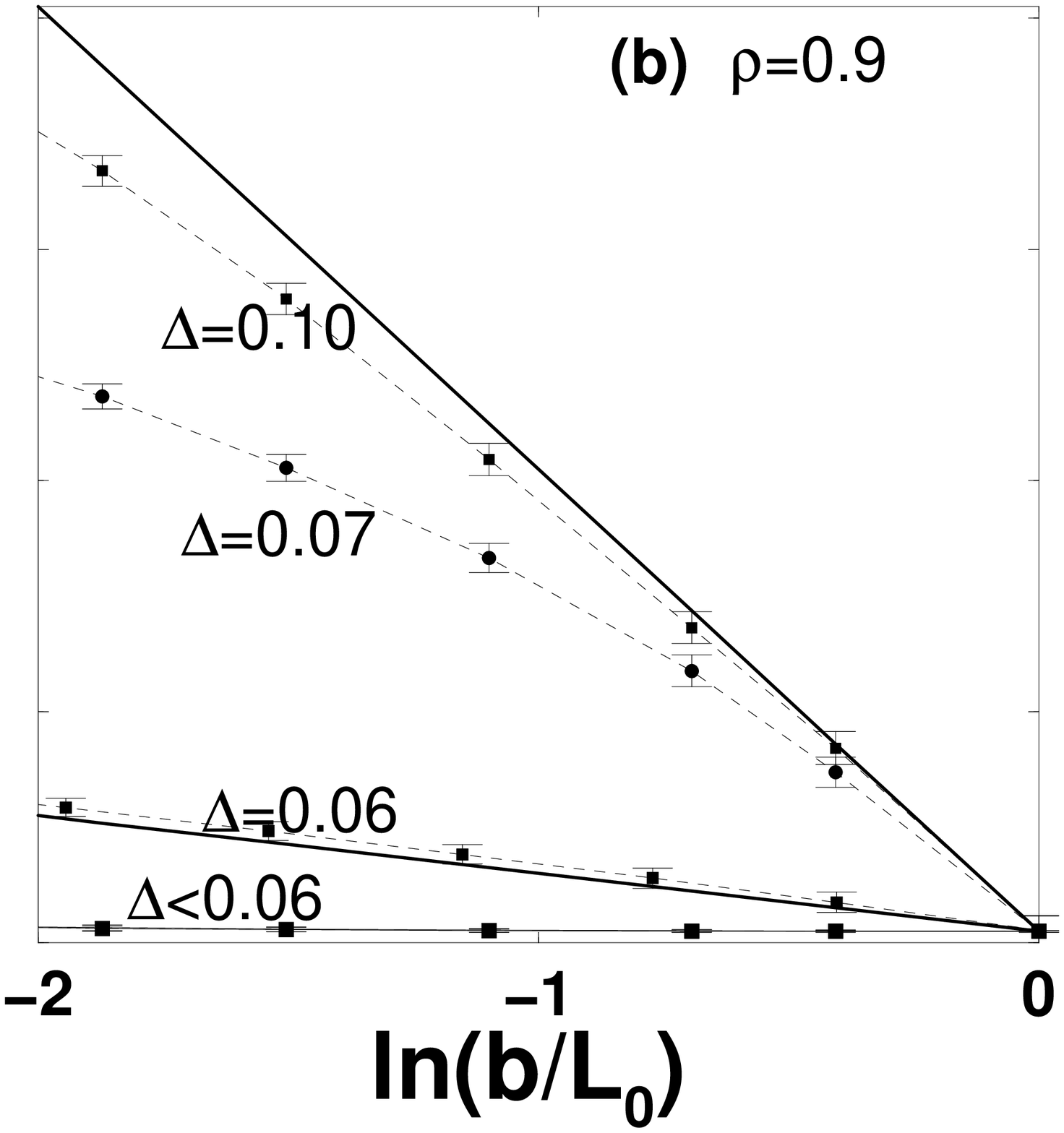}
        \vspace*{0.5cm}
       }
          }     
\caption{Log-log plots of $\langle\psi^2_{b}\rangle$ versus ${b}$ for different
dispersities $\Delta$. Both axes are normalized by their values at system size $L_0$, which
causes the curves to meet at the origin and facilitates comparing their
asymptotic slopes. Dashed lines connecting the data points are guides to
the eye. Solid straight lines are reference lines with slopes $-2$ and
$-1/4$. (a) For $\rho=1.0$ 
there is an abrupt change from asymptotic slope of $0$ (data lying on
the abscissa) to $-2$, which corresponds to a solid-liquid
transition on increasing the dispersity above $\Delta_{th}$.  
(b) For $\rho=0.9$, the $\Delta=0.06$ curve falls between
the solid and liquid regimes, and shows a slope of $-1/4$ characteristic 
of the hexatic phase.}
\label{figpsi2l}
\end{figure}

Next we study the finite size scaling of $\langle\psi^2_{b}\rangle$. 
Because of the qualitative difference in the form of $g_6(r)$ between
solid and liquid, the behavior of $\langle\psi^2_{b}\rangle$ as a
function of ${b}$ changes drastically upon melting
\cite{Bagchi-Andersen}. In the {\em liquid} for ${b}\gg\xi$, where $\xi$
is the correlation length, $\langle\psi^2_{b}\rangle$ decays as
${b}^{-2}$, while for the {\em solid}, $\langle\psi^2_{b}\rangle$
remains constant.  Fig.~\ref{figpsi2l}(a) shows that for $\rho=1.0$, the
behavior of the system changes abruptly from solid (given by the line
with zero slope) to liquid at $\Delta_{th}$, which is consistent
with our previous studies of $g(r)$, $g_6(r)$ and $U_{b}$. The
$\Delta=0.10$ liquid curve shows long-range correlation for ${b}<\xi$,
which crosses over to short-range correlation (slope $-2$) for
${b}\gg\xi$ ($\xi\approx 0.6L_0$ for this curve). The corresponding plots
for larger $\Delta$ show that $\xi$ shrinks upon increasing
$\Delta$. For $\rho=0.9$ (Fig.~\ref{figpsi2l}(b)), we observe both solid
behavior for $\Delta<0.06$ and liquid behavior for $\Delta>0.06$. For
$\Delta=0.06$, Fig.~\ref{figpsi2l}(b) shows an {\em algebraic} decay for
the correlation function, with exponent $-1/4$. This ``intermediate''
behavior is reminiscent of the hexatic phase\cite{Bagchi-Andersen}, for
which the orientational correlation decays algebraically while the
system does not possess quasi-long-range translational order. In
Fig.~\ref{phasediagram} we have specified as the $I$ phase the locus of
the points showing this intermediate behavior.

In summary, we have studied a melting transition driven not by $T$ but
by $\Delta$. We have simulated relatively large systems and applied
finite size scaling (Figs~\ref{figul}, \ref{figpsi2l}) arriving at a
phase diagram for this dispersity-driven melting
(Fig.~\ref{phasediagram}). Melting takes place from a 2D ordered phase
to a disordered liquid phase, similar to the conventional
temperature-driven melting processes. Moreover, at large values of
$\rho$, melting is a first order phase transition at a threshold
dispersity value $\Delta_{th}(\rho)=0.097\pm 0.005$. Our study of the
mean square displacement of the particles shows that this melting is
accompanied by a transition from a frozen solid to a diffusive liquid,
distinguishing it from the glass transition  observed in
\cite{Bocquet}. The threshold line $L_{th}$ extends almost horizontally
down to the point $C$ with coordinates
$\Delta_c\approx0.097,\rho_c\approx0.96$.  Below $\rho_c$, finite size
scaling of the orientational order parameter suggests the existence of
an intermediate ``hexatic'' phase between the solid and liquid
phases. Thus we hypothesize that point $C$ is a multicritical point,
where two lines of continuous transitions (separating liquid/``hexatic''
and ``hexatic''/solid phases) meet the line of first order transitions
$L_{th}$ (separating liquid/solid phases) as shown in
Fig.~\ref{phasediagram}\cite{saito}. Fig.~\ref{phasediagram} provides an
answer to the last two of the three questions: one can {\em not} pass
continuously from solid to liquid ``around the critical point'' $C$,
because the two phases are separated by the line of first order phase
transitions $L_{th}$. A similar horizontal line of order-disorder transition
has been observed in the study of the effect of quenched impurities on the
structure of 2D-solids \cite{Nelson-Fertig}.

We thank N.~Ito for generously introducing us to this topic, S.~T.~Harrington for
significant assistance in the formative stages of this research,
J.~L.~Barrat for bringing Ref.\cite{Bocquet} to our attention,
L.~A.~N.~Amaral, N. Clark, W.~Kob, B.~Kutnjak-Urbanc and D.~R.~Nelson 
%C.~Roscon 
for extremely helpful criticism, NSF for financial support and the Boston University
Center for Computational Science for computational resources.

\end{multicols}

\begin{thebibliography}{99}
\bibitem{simlation} The average simulation speed on Boston University's
SGI power challenge array is $50\mu s$ per particle update, so each of the
$70$ state points studied here requires $16$ hours on one processor
(over $10^3$ hours total computational time).
\end{thebibliography}
\end{document}